\documentclass{PoS}
\usepackage{url}

\def\L{\Lambda}
\def\d{\delta}

\title{Baryons in/and Lattice QCD}

\ShortTitle{Baryons in/and Lattice QCD}

\author{\speaker{Andr\'e Walker-Loud}%
	\thanks{I am indebted to my collaborators for many conversations and contributions to my working understanding of physics and my research.  I am also grateful to my colleagues who were willing to share their often preliminary lattice QCD results for this presentation.  And of course, I am grateful to the organizers for the fun workshop and inviting me to give this presentation.}\\
    Lawrence Berkeley National Laboratory, Berkeley, CA 94720, USA\\
    Department of Physics, University of California, Berkeley, CA 94720, USA \\
    E-mail: \email{awalker-loud@lbl.gov}}

\abstract{
Anchoring our understanding of low-energy nuclear and hadronic physics to the fundamental theory of strong interactions, QCD, remains and outstanding challenge for physicists.
Lattice QCD and chiral perturbation theory are the most powerful theoretical tools we have at our disposal to make this connection.
These tools share a symbiotic relationship as chiral perturbation theory is used to understand the light quark mass dependence of observables, while lattice QCD is used to determine the values of the unknown operator coefficients appearing in the chiral Lagrangian.
In this talk, I review our present understanding of select single baryon properties from lattice QCD and chiral perturbation theory, highlighting some of the challenges and discussing some unresolved puzzles.
}

\FullConference{The 7th International Workshop on Chiral Dynamics,\\
		August 6 -10, 2012\\
		Jefferson Lab, Newport News, Virginia, USA}

\begin{document}

%		Introduction
%%
%%%
%%%%
%%%%%
After three decades of intense effort, lattice QCD (LQCD) has emerged as a robust tool for precisely computing basic low-energy QCD observables.
This development is due to the ever increasing computing power (see eg. \cite{Boyle:2012iy}), and at least as important, the constant development of new computing algorithms (see eg. \cite{Schaefer:2012tq}).
One of the main achievements realized in the last few years is our ability to perform LQCD calculations with the $up$ and $down$ quark masses at or near their physical values.
This heralds a new era in which LQCD will be a principle means of quantitatively understanding low-energy chiral dynamics.

The approximate chiral symmetry of QCD for the light quarks plays a central role in our understanding of hadronic and nuclear physics.
This approximate symmetry is exploited to construct Chiral Perturbation Theory ($\chi$PT)~\cite{chiPT}, which is the low-energy effective field theory~\cite{Weinberg:1978kz} of QCD, and describes the soft pion interactions.
The properties and interactions of these pseudo-scalars are encoded in an infinite tower of operators in the chiral Lagrangian.
The form of these operators are constrained by the symmetries of QCD while the operator coefficients (known as the low-energy constants (LEC)s) are unknown non-perturbative numbers.
Despite the non-renormalizable nature of $\chi$PT, it is still a predictive theory as the quantitative relevance of the various operators are dictated by an expansion of the soft pion momentum and the light quark masses suppressed by the chiral symmetry breaking scale, $\L_\chi\sim 1$~GeV.
Thus, hadronic observables can be computed to a given precision by working to a finite order in the chiral expansion, see eg. the talk of J.~Bijnens at this conference~\cite{Bijnens:2013zc}.
We are now in the era when LQCD can be used to reliably determine the value of these LECs, greatly improving our predictive capabilities and allowing for stringent tests of chiral dynamics and the Standard Model.
Computing the basic properties of the pions and kaons with LQCD has matured to the point that there are now lattice averages for these quantities, see eg. the talk of L.~Lellouch at this conference~\cite{cd:2012}.

While the application of LQCD to properties of the pseudo-scalars has become a precision science (a few percent theoretical uncertainty), there remain significant challenges for computing properties of baryons from LQCD.
These are challenges we must overcome if we are to ultimately make a quantitative connection between the fundamental theory of strong interactions, QCD, and low-energy nuclear physics: the basic properties of the nucleon serve as the first anchor between QCD and our understanding of nuclear physics.
In this talk, I will review the status of our understanding of properties of baryons from LQCD and $\chi$PT.
I will focus on properties of single baryons.  For a review of multiple baryon calculations from LQCD, see the talks of W.~Detmold~\cite{cd:2012} and N.~Ishii~\cite{cd:2012} at this conference and for a good discussion of issues in computing multi-particles correlation functions, see the talk of J.~Dudek~\cite{cd:2012}.

%
%%
%%%
%%%%
%%%%%
\section{Baryons \textit{in} Lattice QCD}

There are two topics I would have liked to cover, but did not for lack of time.
The first topic is LQCD calculations of hadron electromagnetic polarizabilities, discussed by B.~Tiburzi~\cite{Tiburzi:2013um}.
These quantities are particular sensitive to chiral dynamics and will really put $\chi$PT to the test; at LO in the chiral expansion, the polarizabilities scale as $\alpha_E,\beta_M \sim 1/m_\pi \sim 1/\sqrt{m_q}$, exposing the predicted non-analytic light quark mass dependence.
Further, at present, there is a discrepancy between the two-loop $\chi$PT prediction of the pion polarizabilites and the measured polarizability, see eg. the talk of D.~Lawrence~\cite{cd:2012}. 
There are also significant uncertainties in the nucleon polarizabilities, particularly that of the neutron (for a recent review, see \cite{Griesshammer:2012we}), and as I will discuss at the end of my talk, the uncertainty of iso-vector nucleon magnetic polarizability is currently the dominant uncertainty in determining the electromagnetic self-energy contribution to $m_p - m_n$~\cite{WalkerLoud:2012bg}.
For these reasons there are several efforts to compute hadron polarizabilities from LQCD, see eg.~\cite{LQCD:POLAR}.

The second topic I would have liked to discuss are the scalar quark matrix elements of the nucleon, $m_q\,  \langle N | \bar{q}\, q | N \rangle$; discussed at this conference by J.~Martin-Camalich~\cite{Camalich:2013zqa}.
These matrix elements are important for understanding spin-independent direct dark matter searches~\cite{qbarq}, yet they are challenging to determine experimentally.
LQCD is a perfect tool for this problem, and the status of such calculations was reviewed at the most recent Lattice Field Theory Symposium~\cite{Young:2013nn}.
There is only one lattice calculation so far with control of all the systematics in the calculations~\cite{Durr:2011mp}, but still with larger than desired uncertainties.
A lattice average of $m_s\,  \langle N | \bar{s}\, s | N \rangle$ was recently performed finding reassuring agreement between several different lattice calculations, but also a smaller value than previously thought~\cite{Junnarkar:2013ac}.

%		Matrix Elements
%%
%%%
%%%%
%%%%%
\subsection{Nucleon matrix elements}

I begin this topic by referring to the talk of D.~Renner~\cite{cd:2012}; I agree with his conclusions which I paraphrase "\textit{For baryon matrix elements, lattice calculations currently lack sufficient study of systematic effects: finite volume, excited state contamination, continuum limit, ... Apparent conflicts with experimental measurements are not justified and apparent conflicts with $\chi$PT are not compelling either}".
If we do not take these cautions seriously, then we are forced to ask \textit{Is there something wrong with QCD?} or \textit{Is there something wrong with our LQCD calculations?}

%		gA
%%
%%%
%%%%
%%%%%
\subsubsection{The nucleon axial charge and other nucleon matrix elements}
%%%%%%%%%%%%%%%%%%%%%%%%%%%%%%%%%%%%%%%%%
%%%  FIG: gA
%%%%%%%%%%%%%%%%%%%%%%%%%%%%%%%%%%%%%%%%%
\begin{figure}[b]
\center
\begin{tabular}{cc}
\includegraphics[width=0.4\textwidth]{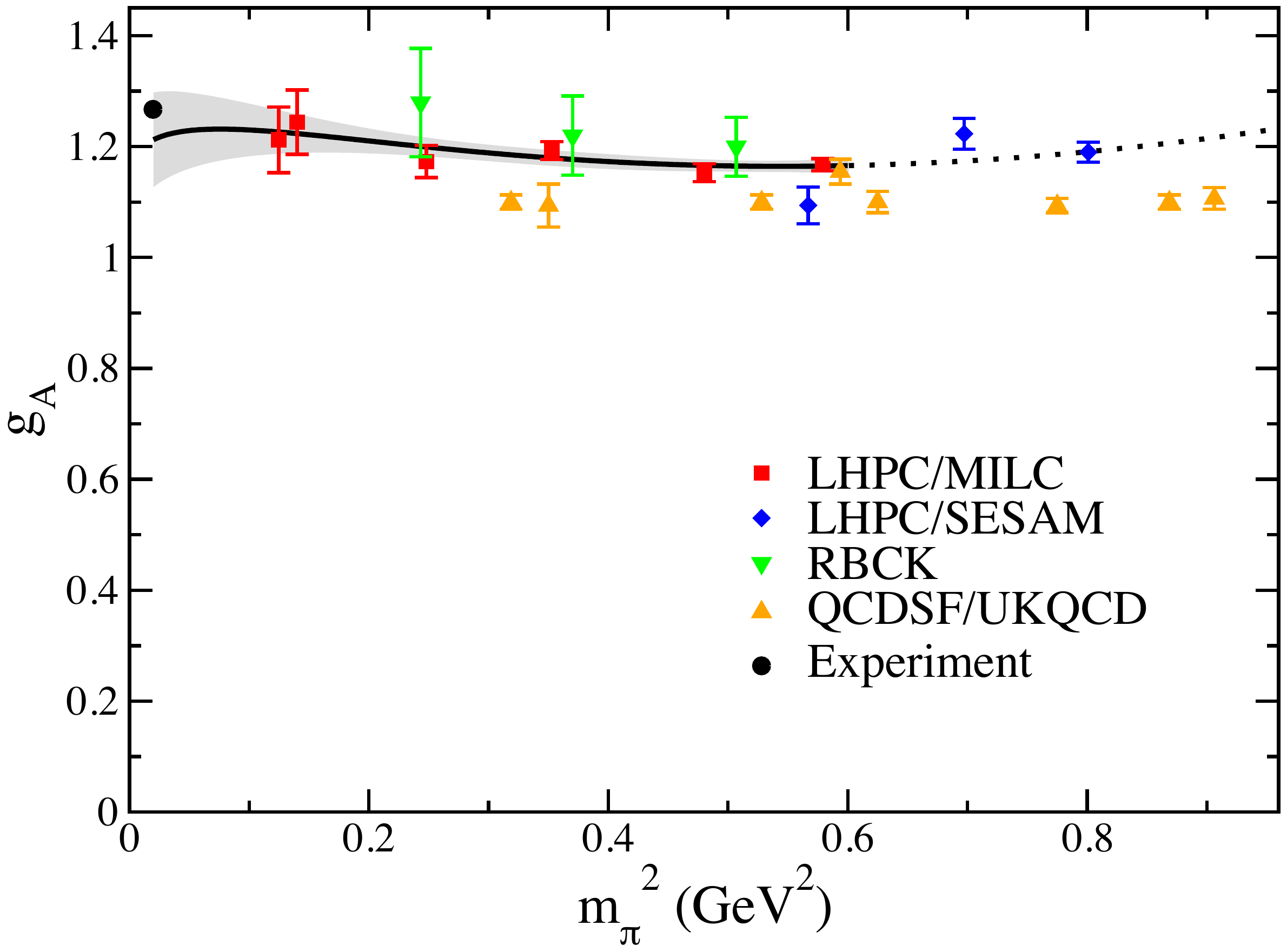}&
\includegraphics[width=0.45\textwidth]{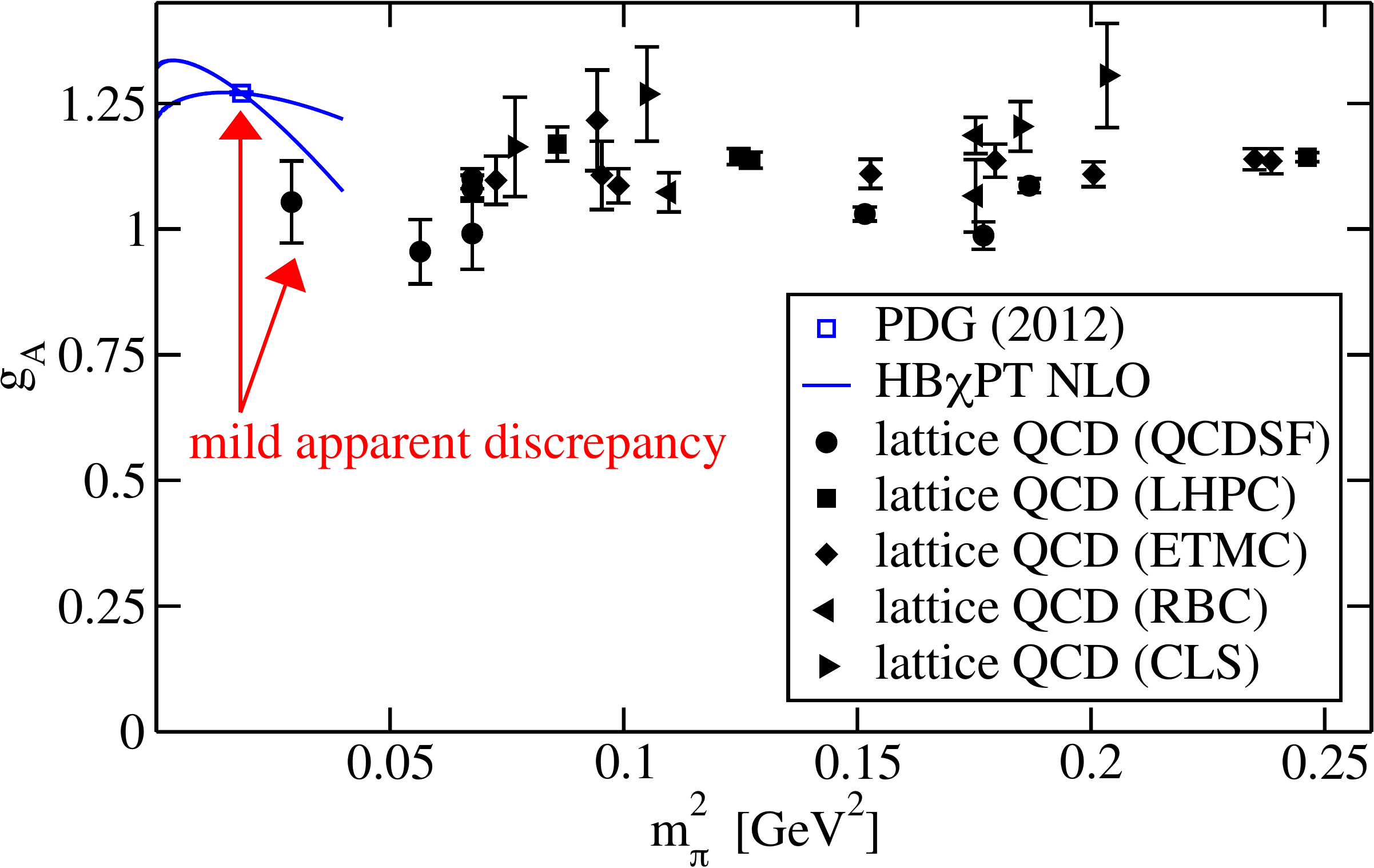}
\end{tabular}
\caption{\label{fig:gA}Left figure from LHPC~\cite{Edwards:2005ym}.  Right figure, summary of lattice results from Dru Renner~\cite{cd:2012}.}
\end{figure}
%%%%%%%%%%%%%%%%%%%%%%%%%%%%%%%%%%%%%%%%%
The nucleon axial charge, $g_A$, has proven particularly challenging to reproduce with LQCD calculations.
In 2005, a promising calculation of $g_A$ was performed with the lightest pion mass of $m_\pi \sim 300$~MeV and a chiral extrapolation that agreed with the experimental value~\cite{Edwards:2005ym}.
However, more recent calculations with lighter pion masses are all further from the experimental value, in disagreement with the chiral extrapolation of LHPC~\cite{Edwards:2005ym}, see Figure~\ref{fig:gA}.

One notices an absence of significant pion mass dependence.
There is clearly some cancelation between different orders in the chiral expansion, but how severe is the cancelation?
In the large $N_c$ expansion, the axial coupling to the nucleon scales as $N_c$.
It turns out, the one-loop corrections to $g_A$ are suppressed in the large $N_c$ expansion~\cite{gA_largeN} and so the size of the higher order chiral corrections are not as large as one would otherwise guess.  The cancellation is only apparent when one explicitly includes the delta degrees of freedom~\cite{FloresMendieta:2000mz}; baryon $\chi$PT calculations which do not explicitly include the delta degrees of freedom will miss these and other important quantitative effects and result in \textit{unnatural} value for LECs.

\begin{figure}
\center
\begin{tabular}{cc}
\includegraphics[width=0.4\textwidth]{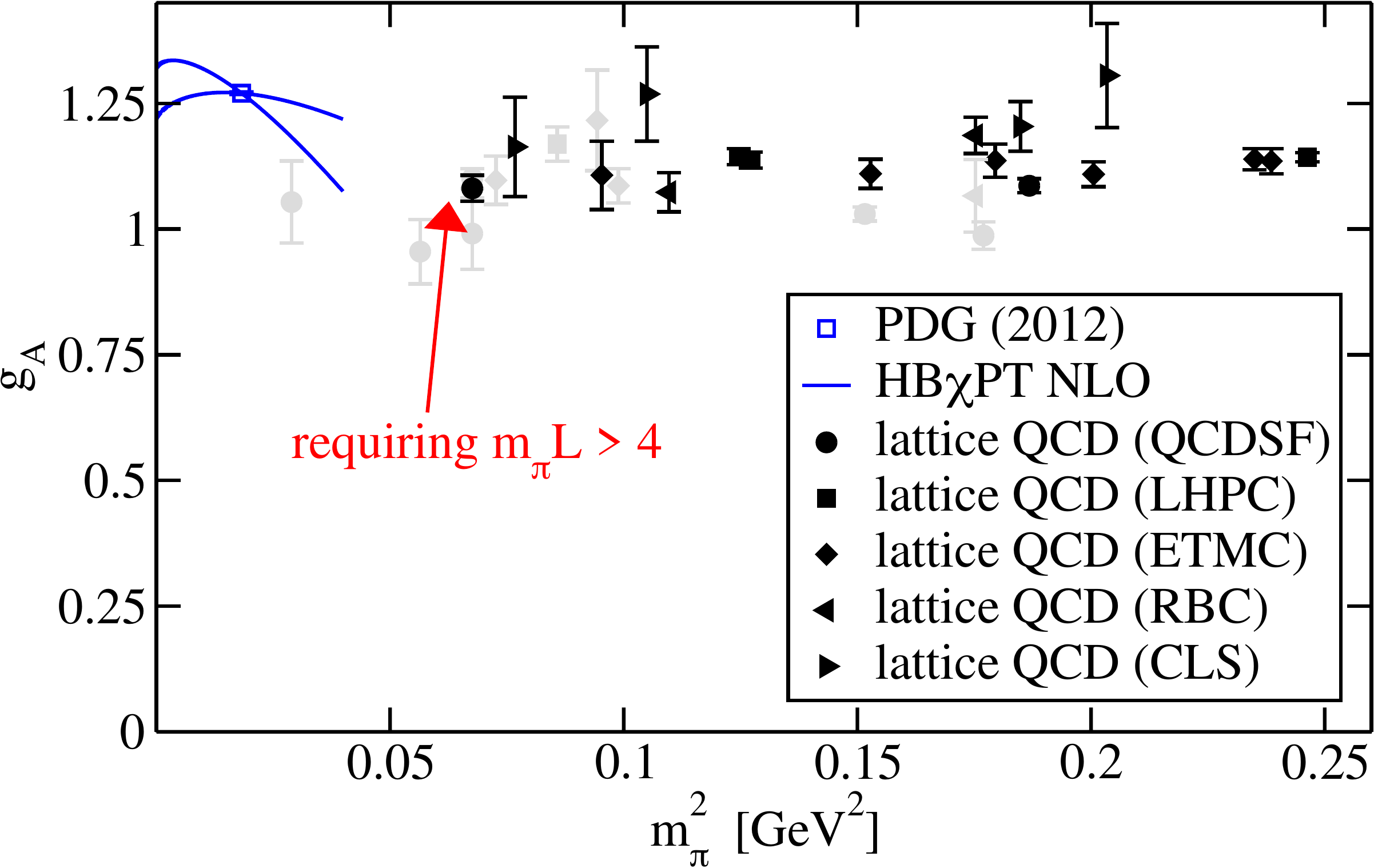}&
\includegraphics[width=0.4\textwidth]{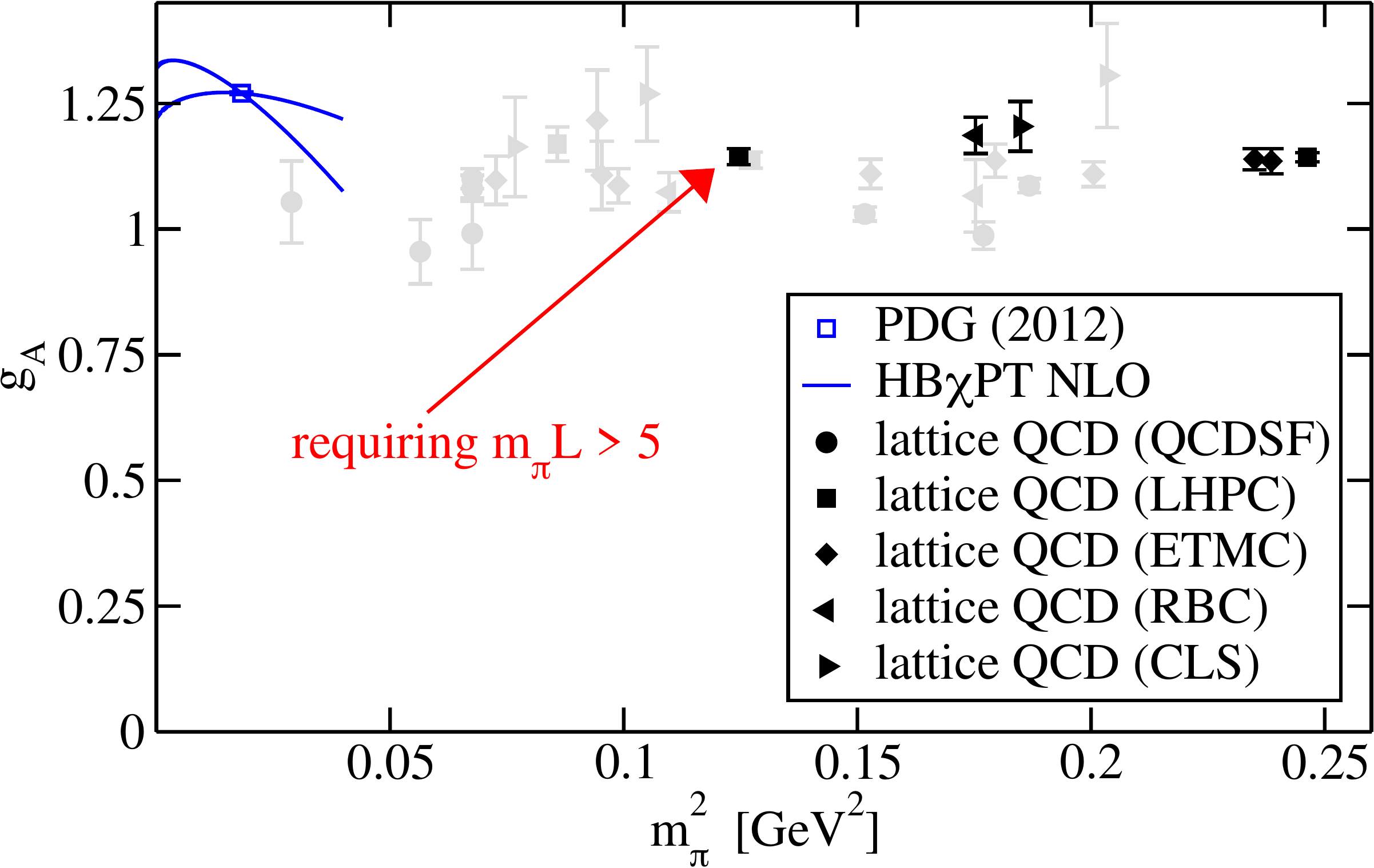}
\\
$m_\pi L \geq 4$& $m_\pi L \geq 5$
\end{tabular}
\caption{\label{fig:gA_cut}  Lattice calculations (courtesy of Dru Renner~\cite{cd:2012}) with cuts finite volume cuts applied.
}
\end{figure}

In the light pion mass region (defined here as $m_\pi \lesssim 300$~MeV) the calculated values of $g_A$ tend to decrease while the physical value is larger than the computed values.  This has caused some alarm~\cite{Lin:2012ev}.
A rule of thumb that is often invoked is that $m_\pi L \geq 4$ for confidence that finite volume modifications are a few percent or less.
If we simply apply this cut to the lattice results for $g_A$, we arrive at the left figure in Figure~\ref{fig:gA_cut}.
%%%%%%%%%%%%%%%%%%%%%%%%%%%%%%%%%%%%%%%%%
%%%  FIG: gA mpi L cut
%%%%%%%%%%%%%%%%%%%%%%%%%%%%%%%%%%%%%%%%%
%%%%%%%%%%%%%%%%%%%%%%%%%%%%%%%%%%%%%%%%%
With baryons, there are reasons to believe $m_\pi L \geq 5$ may be needed.
Applying these cuts removes the apparent discrepancy with the experimental value, but unfortunately also removes most of the numerical results.

\begin{figure}
\center
\begin{tabular}{cc}
\includegraphics[width=0.42\textwidth]{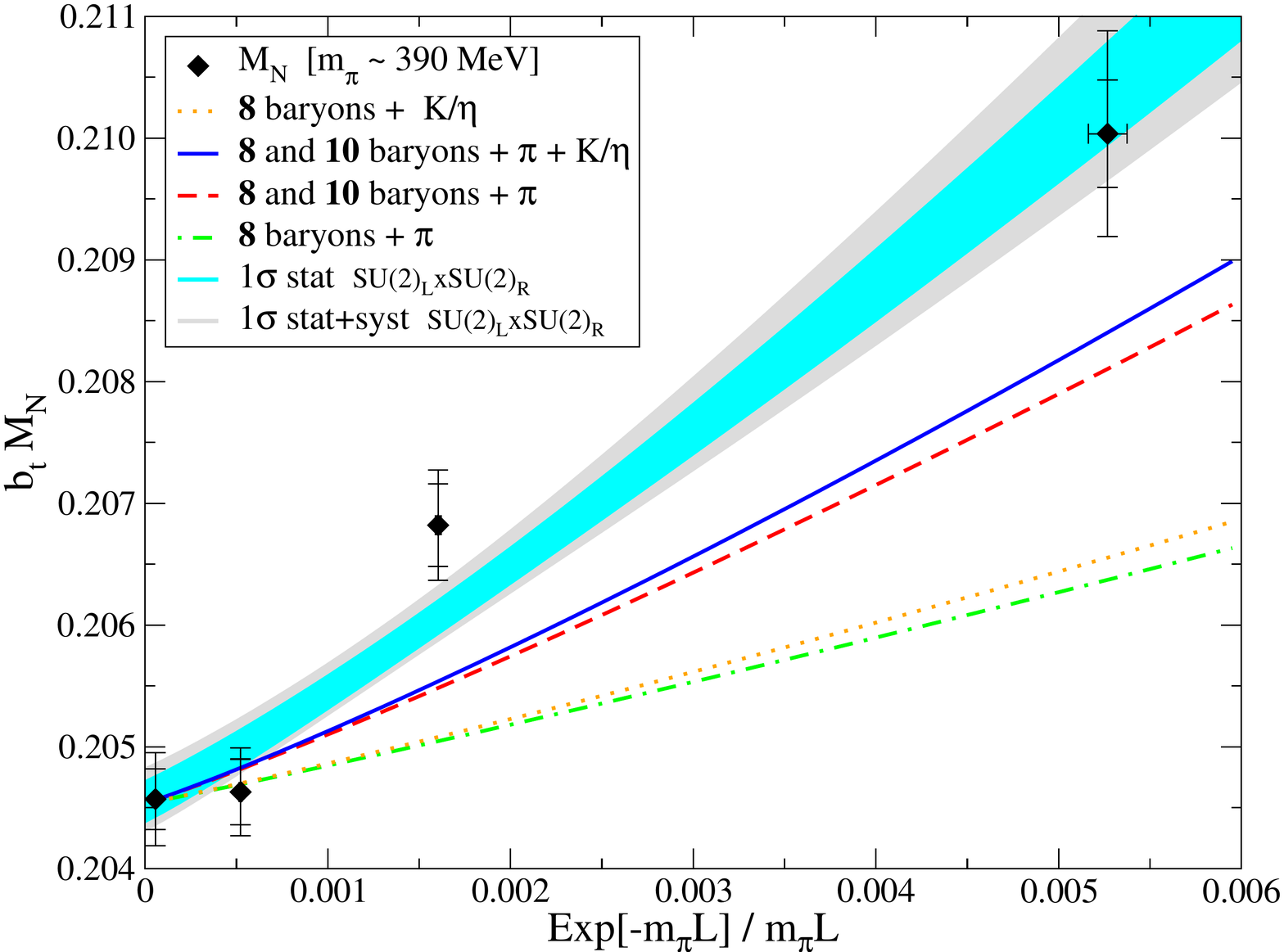}&
\includegraphics[width=0.42\textwidth]{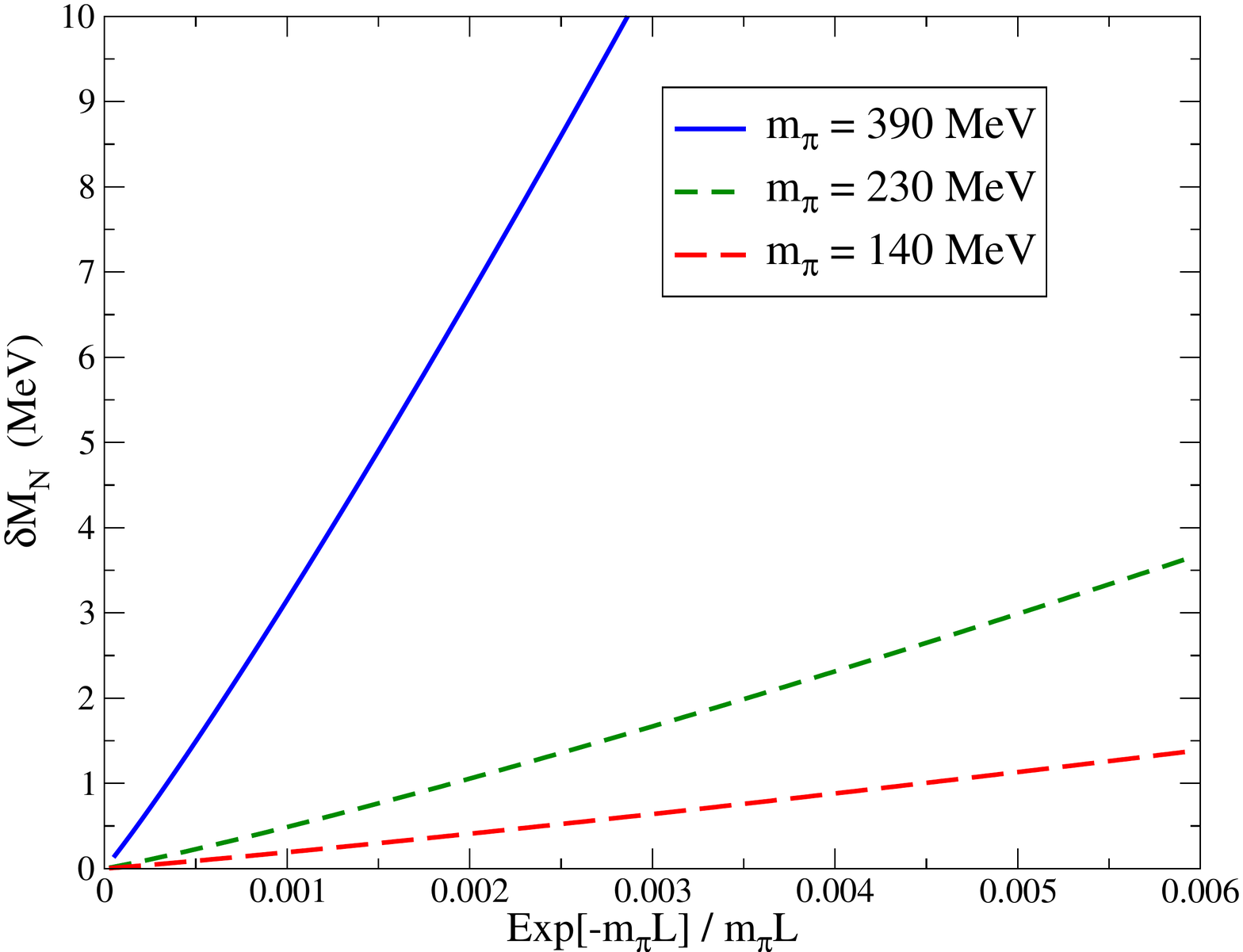}
\end{tabular}
\caption{\label{fig:mN_FV}  Finite volume dependence of the nucleon from Ref.~\cite{Beane:2011pc}.
}
\end{figure}

As an example, I display two figures from Ref.~\cite{Beane:2011pc}, which computed the single hadron spectrum at $m_\pi \sim 390$~MeV, Figure~\ref{fig:mN_FV}.
The left figure displays the computed nucleon mass (in lattice units) on four different volumes.  From left to right (largest to smallest) these volumes satisfy $m_\pi L \simeq \{7.7, 5.8, 4.8, 3.9 \}$.
Only the largest two volumes have negligible finite volume corrections and with $m_\pi L \simeq 3.9$ we observed a 2.5\% correction, which is more significant than expectations based purely on baryon $\chi$PT~\cite{Beane:2004tw}.
In the large volume limit, the finite volume corrections are approximated by (ignoring explicit $\Delta$ loops)
\begin{equation}
\label{eq:mn_FV}
\Delta m_N^{FV} = \frac{3\pi g_A^2 m_\pi^3}{(4\pi f_\pi)^2}
	\sum_{\vec{n} \neq 0} \frac{e^{-|\vec{n}| m_\pi L}}{|\vec{n}|m_\pi L}\, ,
\end{equation}
(with $f_\pi \simeq 131$~MeV at the physical pion mass).
For fixed $m_\pi L$, the volume corrections are predicted to scale roughly as $m_\pi^3$.
The right plot of Figure~\ref{fig:mN_FV} displays the estimated volume dependence for different pion masses.
Similarly, the volume corrections for $g_A$ are approximately
\begin{equation}
\label{eq:gA_FV}
\Delta g_A^{FV} = -g_A \frac{m_\pi^2}{(4\pi f_\pi)^2}
	\sum_{\vec{n} \neq 0} e^{-|\vec{n}| m_\pi L} \sqrt{\frac{\pi}{2|\vec{n}| m_\pi L}} \left(
		\frac{8 +6 g_A^2}{|\vec{n}| m_\pi L}
		-\frac{16g_A^2}{3}
	\right)\, :
\end{equation}
for fixed $m_\pi L$, they are expected to scale roughly as $m_\pi^2$.
Care must be taken when assessing the expected importance of finite volume corrections.  Often cuts are made based on $m_\pi L$, but this can be too simplistic as the  corrections also have non-negligible pion mass dependence.
%%%%%%%%%%%%%%%%%%%%%%%%%%%%%%%%%%%%%%%%%
%%%  FIG: mB volume dependence
%%%%%%%%%%%%%%%%%%%%%%%%%%%%%%%%%%%%%%%%%
%%%%%%%%%%%%%%%%%%%%%%%%%%%%%%%%%%%%%%%%%

\begin{figure}[b]
\center
\begin{tabular}{cc}
\includegraphics[width=0.4\textwidth]{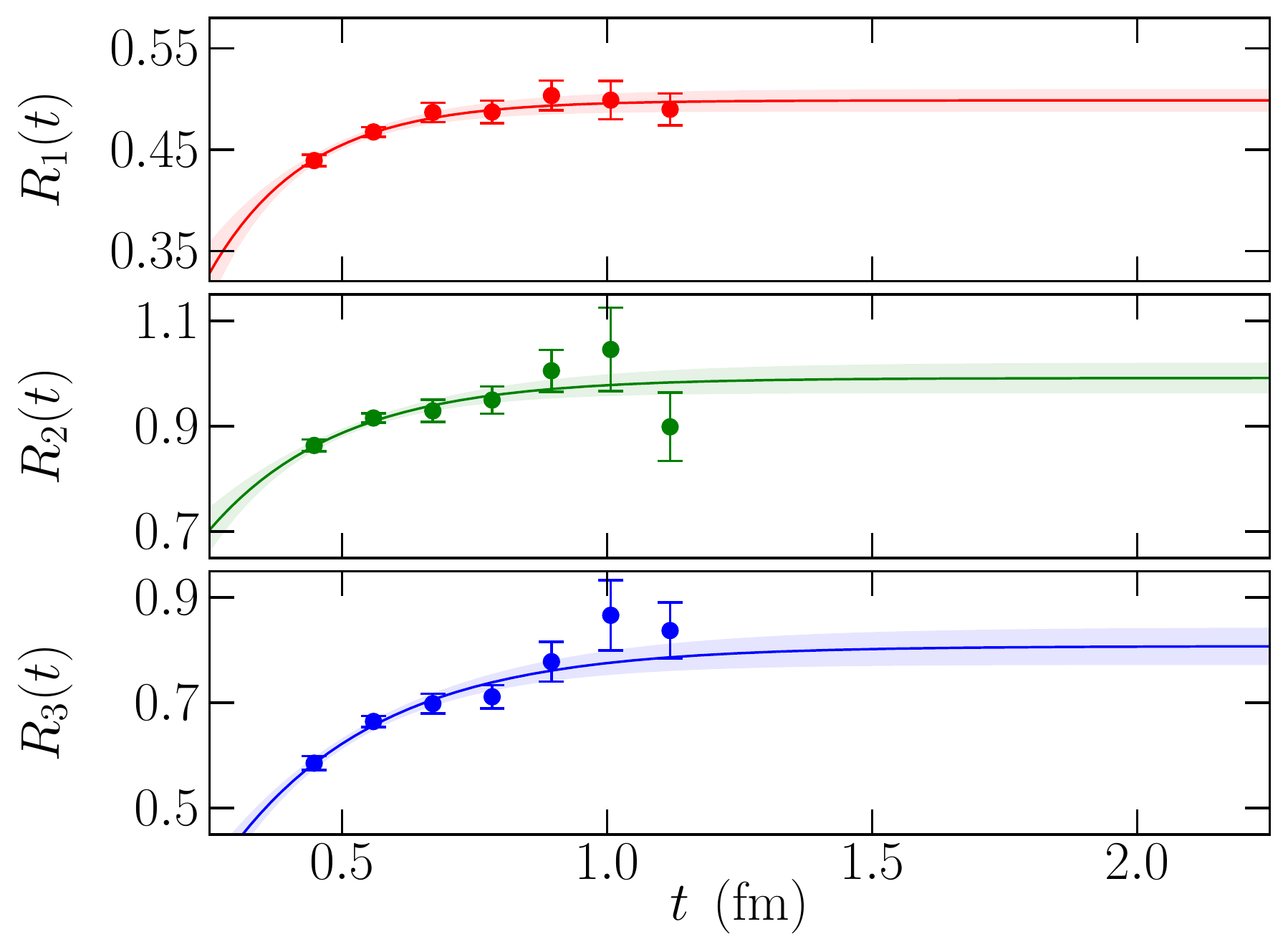}&
\includegraphics[width=0.45\textwidth]{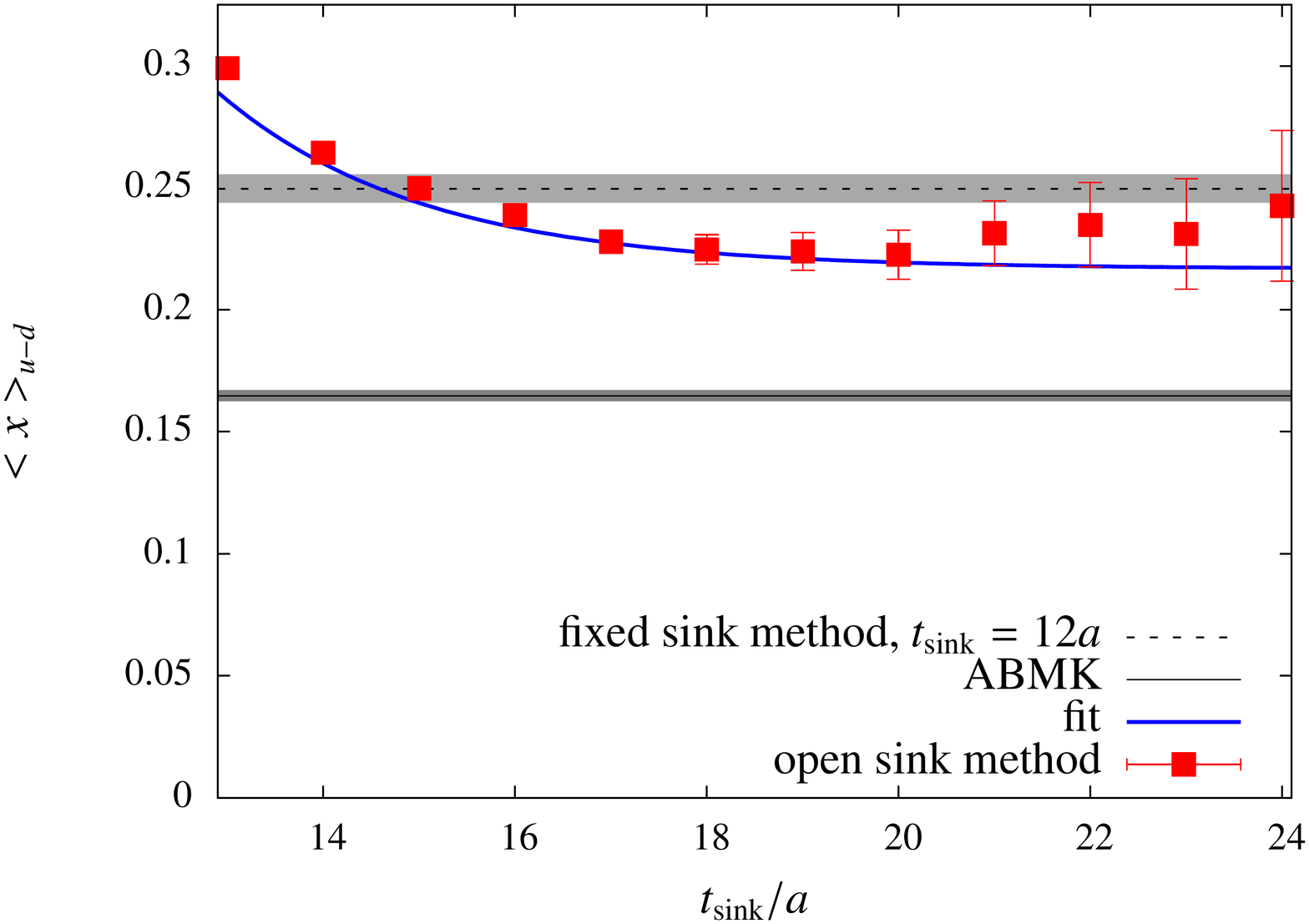}
\end{tabular}
\caption{\label{fig:matrix_elements}  Two techniques for controlling excited state contamination: \cite{Meinel_axial} (left) and \cite{Dinter:2011sg} (right).
}
\end{figure}

Finite volume corrections are one of a few systematics that we must control in LQCD calculations.
Other systematics are: the chiral extrapolation, the continuum extrapolation, renormalization when necessary, and contamination from excited states.
The chiral, volume and continuum systematics have been most thoroughly explored in lattice calculations.  Many groups are now focussing on exploring the contamination from excited states.
Here, I highlight two different approaches, displayed in Figure~\ref{fig:matrix_elements}.
%%%%%%%%%%%%%%%%%%%%%%%%%%%%%%%%%%%%%%%%%
%%%  FIG: New Matrix Element Methods
%%%%%%%%%%%%%%%%%%%%%%%%%%%%%%%%%%%%%%%%%
%%%%%%%%%%%%%%%%%%%%%%%%%%%%%%%%%%%%%%%%%
The matrix elements are typically constructed with three-point functions
\begin{equation}\label{eq:3pt}
C_{3pt}(t,t^\prime) = \sum_{\vec{x}} \sum_{\vec{y}} 
	e^{-i \vec{p} \cdot \vec{x}} e^{i \vec{q} \cdot \vec{y}}
	\langle 0| H(\vec{x},t)\, O_g(\vec{y},t^\prime)\, H^\dagger(\vec{0},0) | 0 \rangle
\end{equation}
where $H^\dagger(\vec{x},t)$ creates hadrons with given quantum numbers, the operator $O_g(\vec{y},t^\prime)$ is related to the matrix element $g$ of interest and the Fourier transforms project onto definite momentum.
For matrix element calculations, using the ``standard ratio" method, where the sink time $t$ is fixed while $t^\prime$ is varried, one can show expected contamination from excited states is given approximately as
\begin{equation}\label{eq:rt}
R(t) = g + \delta_Z e^{-\delta_E t} + \dots
\end{equation}
where $t$ is the time separation between the source and sink with the operator insertion time  chosen as $t^\prime = t/2$,
$g$ is the value of the matrix element of the state of interest,
$\delta_Z$ is related to ratios of overlap factors of the interpolating fields onto the state of interest and the excited state 
and $\delta_E = E_{exc} - E_0$ is the gap between the state of interest and the first excited state.
In \cite{Meinel_axial}, the heavy-hadron axial couplings were computed for several values of the source-sink separation time and extrapolated to infinite separation using Eq.~(\ref{eq:rt}); the results are displayed in the left plot of Figure~\ref{fig:matrix_elements}.

Another method being studied is the "open sink" method~\cite{Dinter:2011sg} where the operator insertion time $t^\prime$ is fixed and the sink time $t$ is left free, Eq.~(\ref{eq:3pt}).  
In the right panel of Figure~\ref{fig:matrix_elements}, this method is applied to the iso-vector momentum fraction $\langle x \rangle_{u-d}$ in the nucleon.
Again, one observes clear contamination from excited states affecting the ``standard ratio" method, which are quantified and removed with this new technique.
These methods have not resolved the nucleon $g_A$ puzzle.

Another promising technique~\cite{3pt_sum} has very recently been utilized in computations of the nucleon electromagnetic form factors.
The calculation was performed by LHP~\cite{Green:2012ud} with a lattice spacing of $a\sim 0.116$~fm, a spatial volume of $L\sim 4.3$~fm and excitingly a pion mass very near its physical value, $m_\pi \simeq 150$~MeV.
%%%%%%%%%%%%%%%%%%%%%%%%%%%%%%%%%%%%%%%%%
%%%  FIG: Form Factors
%%%%%%%%%%%%%%%%%%%%%%%%%%%%%%%%%%%%%%%%%
\begin{figure}[b]
\center
\begin{tabular}{cc}
\includegraphics[width=0.45\textwidth]{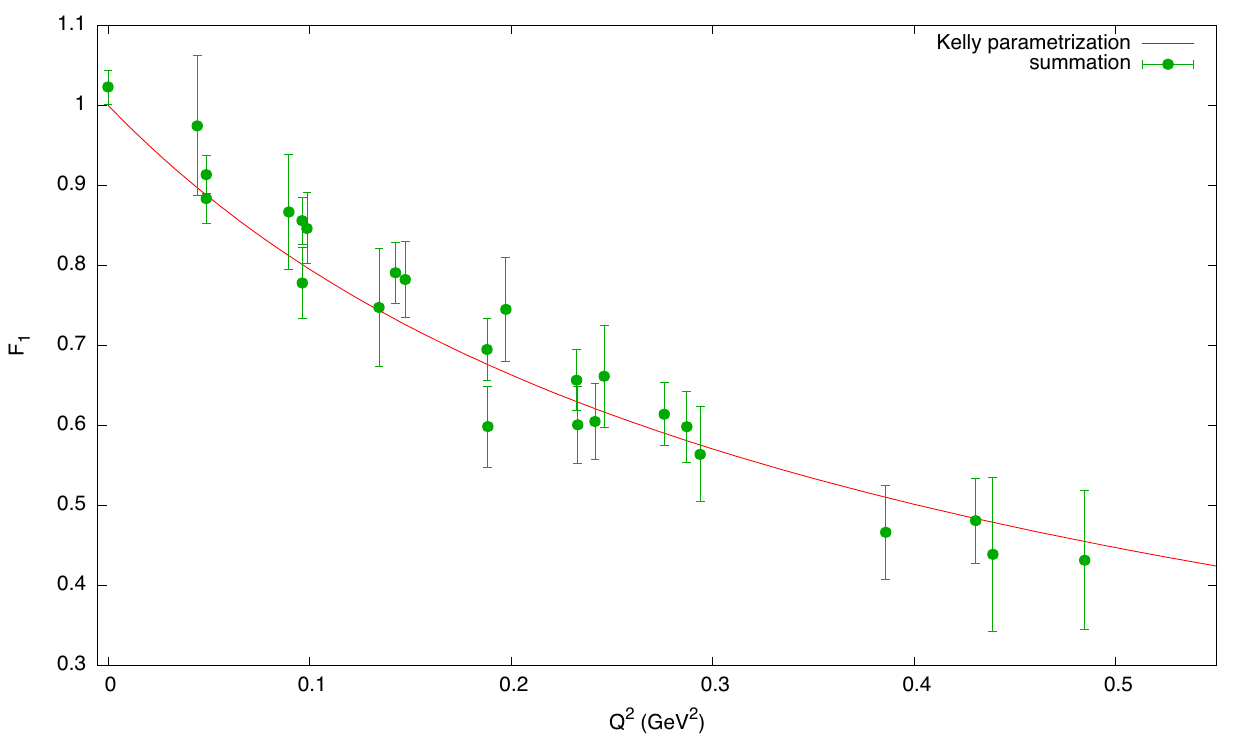}&
\includegraphics[width=0.45\textwidth]{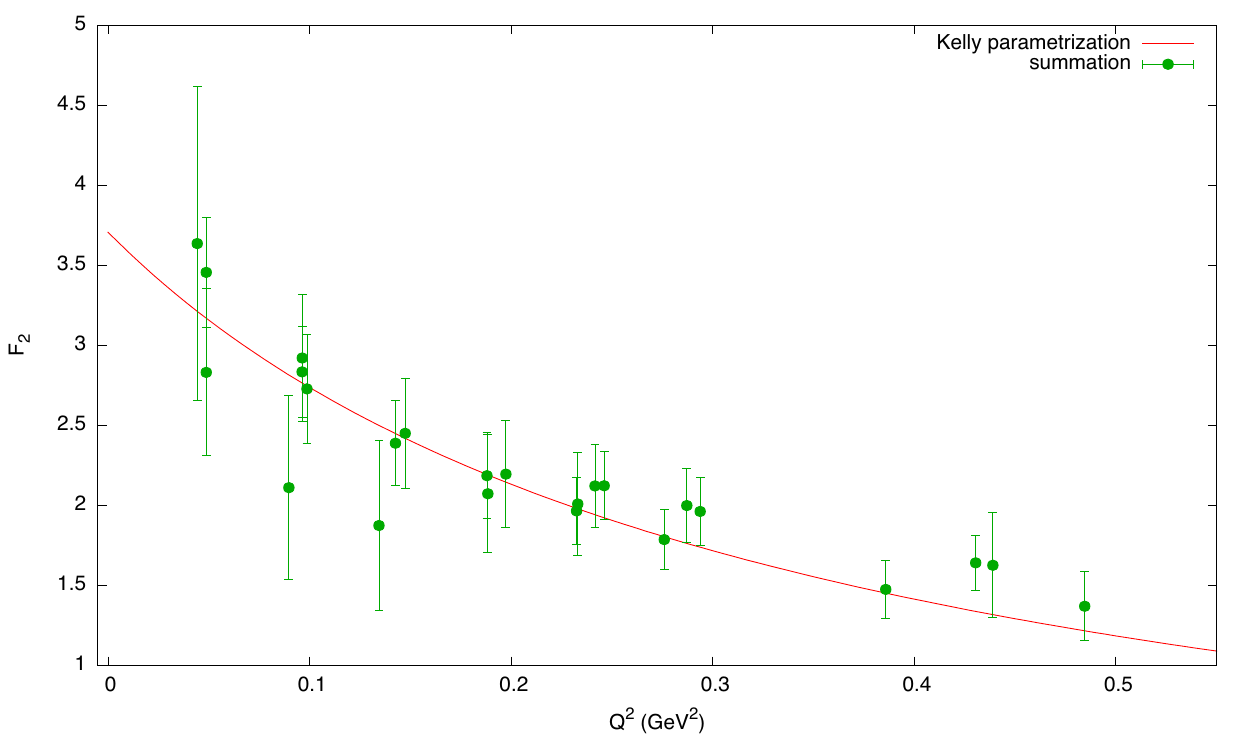}
\end{tabular}
\caption{\label{fig:form_factors}  Calculation of Dirac and Pauli form factors of the nucleon with nearly physical pion masses~\cite{Green:2012ud}.
}
\end{figure}
%%%%%%%%%%%%%%%%%%%%%%%%%%%%%%%%%%%%%%%%%
What is particularly exciting about the results is that for the first time, the resulting form factors are in remarkable agreement with the experimentally measured form factors.
Figure~\ref{fig:form_factors} displays the resulting lattice calculations for the $F_1(Q^2)$ (left) and $F_2(Q^2)$ (right) elastic iso-vector nucleon form factors (green points).  Additionally, the Kelly parameterization of the experimentally measure form factors is displayed as the solid (red) curve.
Unfortunately, this method did not resolve the $g_A$ puzzle.

In my opinion, the most promising approach for computing these and other matrix elements is to use a variational method.  With this approach, one constructs a basis of interpolating fields that are diagonalized to project onto the various eigenstates of QCD.
This technique has been applied very successfully to the ground and excited state spectrum of QCD, see for eg. the talk of J.~Dudek at this conference~\cite{cd:2012}.
The success arises partly because the variational diagonalization of the interpolating fields allows for an extraction of the matrix elements much earlier in Euclidean time before the signal-to-noise has decayed.  See Ref.~\cite{Owen:2012ts} for an example of this method applied to a calculation of $g_A$.  It was found excited states can systematically suppress the computed value by as much as 12\%.

%		mq dependence of mB
%%
%%%
%%%%
%%%%%
\subsection{Light quark mass dependence of the nucleon (and other baryons)}

\begin{figure}
\center
\begin{tabular}{cc}
\includegraphics[width=0.45\textwidth]{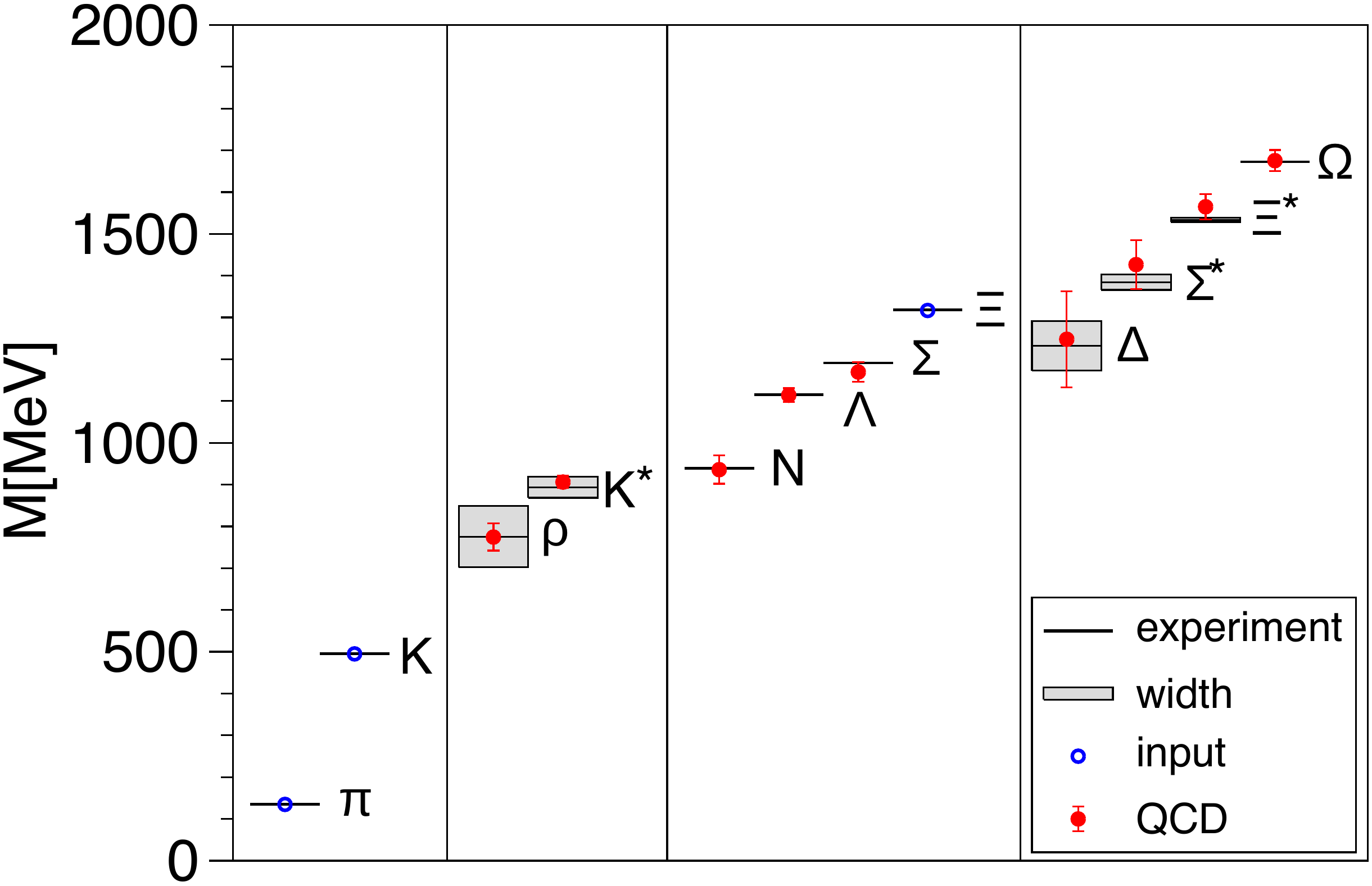}&
\includegraphics[width=0.42\textwidth]{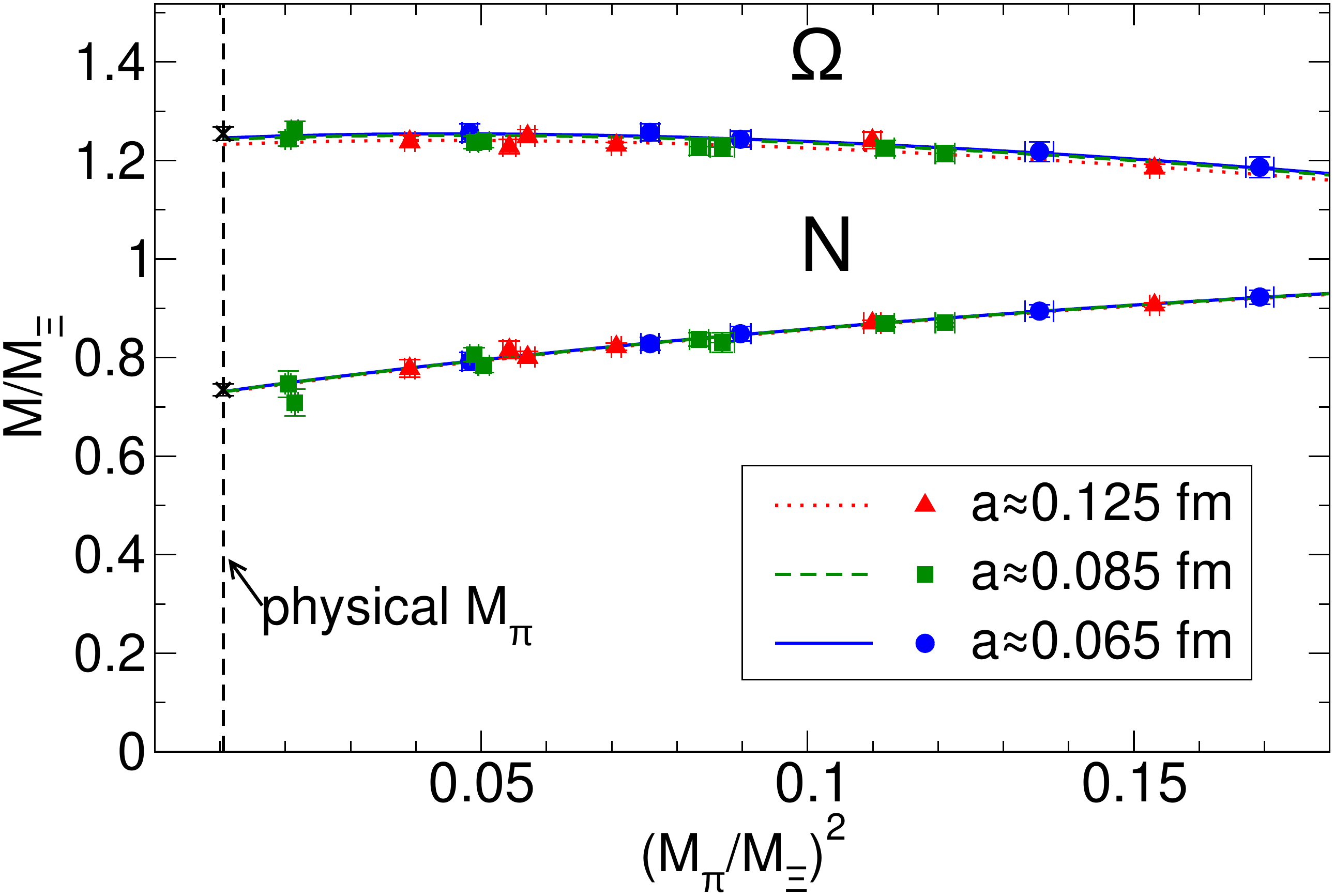}
\end{tabular}
\caption{\label{fig:bmw_spec}  Ground state hadron spectrum calculation and sample chiral extrapolation of Ref.~\cite{Durr:2008zz}.
}
\end{figure}

I would like to transition to a discussion of the ground state baryon spectrum and some unresolved puzzles concerning their light quark mass dependence.
In 2008, the Budapest-Marseille-Wuppertal Collaboration presented results of the ground state hadron spectrum (subsequently published in Science~\cite{Durr:2008zz}) composed of $u, d$ and $s$ quarks at the XXVI International Symposium on Lattice Field Theory, held up the road in Williamsburg.
For the first time, all systematics were sufficiently controlled so as to make a meaningful quantitative comparison with the spectrum of nature:
the calculations were performed with $m_\pi \gtrsim 185$~MeV, three lattice spacings and relatively large volumes, resulting in extrapolated results that agree with the experimental (isospin averaged) masses within a few percent, see Figure~\ref{fig:bmw_spec}.
This was an exciting development that took at least some of us by surprise.
%%%%%%%%%%%%%%%%%%%%%%%%%%%%%%%%%%%%%%%%%
%%%  FIG: Form Factors
%%%%%%%%%%%%%%%%%%%%%%%%%%%%%%%%%%%%%%%%%
%%%%%%%%%%%%%%%%%%%%%%%%%%%%%%%%%%%%%%%%%

At the same conference, an interesting puzzle concerning the light quark mass dependence of the nucleon~\cite{WalkerLoud:2008bp} was brought to light; in all LQCD calculations with dynamical light and strange quarks (including \cite{Durr:2008zz}), the nucleon mass displays a remarkably linear dependence on the pion mass~\cite{WalkerLoud:2008pj}, now known as the \textit{ruler approximation}.
The lattice results are also consistent with a fit to the baryon $\chi$PT formula, albeit with large cancelations between different orders in the chiral expansion, see Refs.~\cite{WalkerLoud:2008bp,WalkerLoud:2008pj} for more details.
A detailed $SU(2)$ baryon $\chi$PT analysis led to the conclusion that the perturbative expansion for the nucleon mass converges only in the range $m_\pi \lesssim 350$~MeV, a conclusion first made in Ref.~\cite{Beane:2004ks}.
The resulting extrapolations gave
\begin{equation}
m_N = \left\{ \begin{array}{lll}
	941\pm42\pm17& \textrm{MeV,}\quad& \textrm{NNLO $SU(2)$ baryon $\chi$PT} \\
	938\pm9 &\textrm{MeV,}\quad& m_N = \alpha_0 + \alpha_1 m_\pi \textit{ (ruler approximation)}
	\end{array}
	\right.
	\, .
\end{equation}

\begin{figure}
\center
\begin{tabular}{cc}
\includegraphics[width=0.45\textwidth]{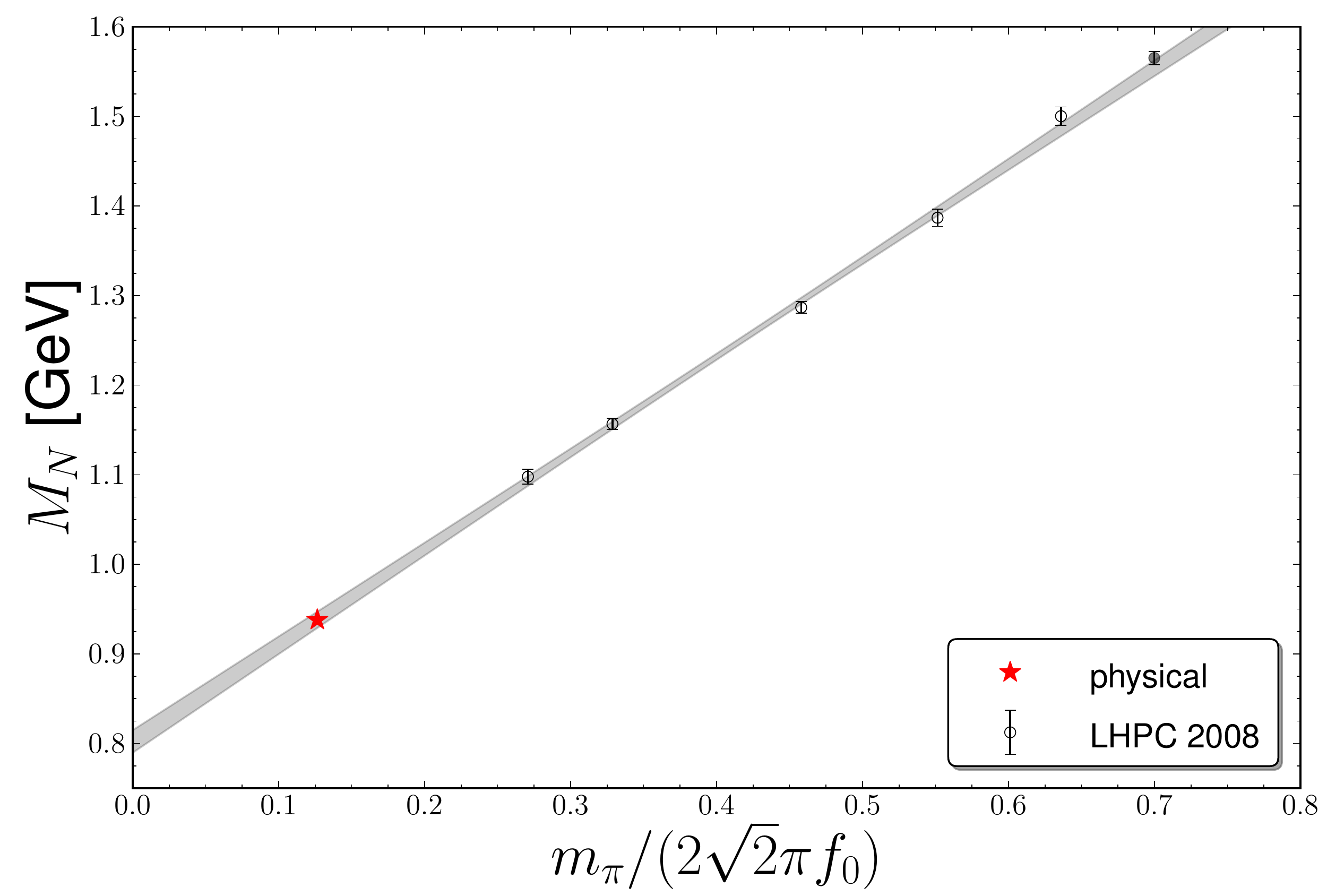}&
\includegraphics[width=0.45\textwidth]{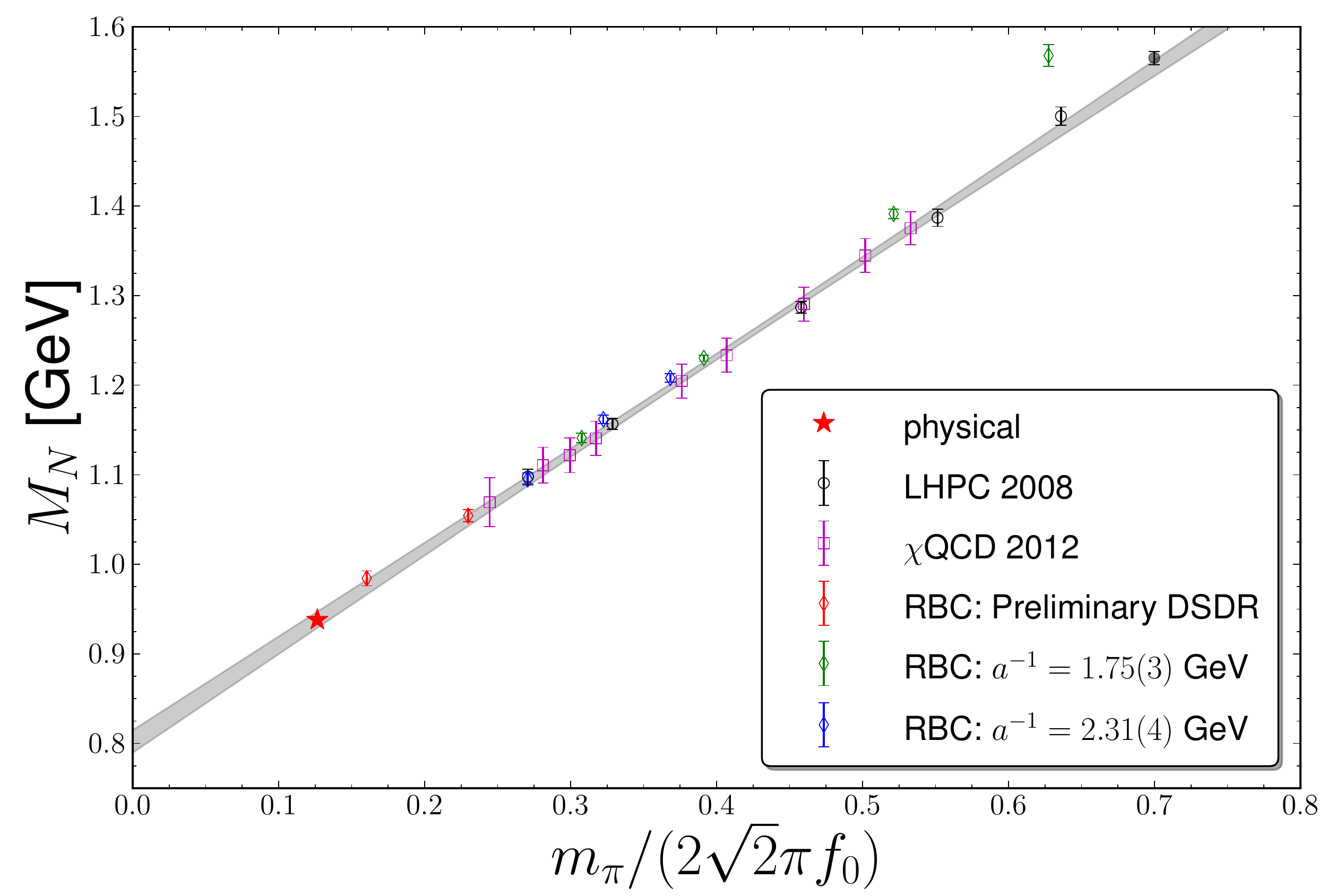}
\end{tabular}
\caption{\label{fig:mN}  Nucleon mass versus pion mass and \textit{ruler approximation}, $m_N = \alpha_0 + \alpha_1 m_\pi$.
Left panel: original fit from Ref.~\cite{WalkerLoud:2008bp} (physical point not included in the fit).  Right panel: new results overlayed with the results of \cite{WalkerLoud:2008bp}.
Within uncertainties, the nucleon mass is given by $m_N[\textrm{MeV}] = 800 + m_\pi$.
}
\end{figure}

What is the status now?
In Figure~\ref{fig:mN}, the \textit{ruler approximation} of Refs.~\cite{WalkerLoud:2008bp,WalkerLoud:2008pj} is displayed in the left panel.  The physical value (red-star) is not included in the fit.
The $x$-axis has been scaled by $2\sqrt{2}\pi f_0\simeq 1083$~MeV such that it is approximately the dimensionless expansion parameter relevant for baryon $\chi$PT.
In the right panel, the most recent results from RBC-UKQCD (unpublished) as well as those from $\chi$QCD~\cite{Gong:2013vja} are overlayed on the original LHP results.  The linear pion mass dependence is observed to persist in the LQCD results over the range $170 \lesssim m_\pi \lesssim 760$~MeV!
%%%%%%%%%%%%%%%%%%%%%%%%%%%%%%%%%%%%%%%%%
%%%  FIG: mN
%%%%%%%%%%%%%%%%%%%%%%%%%%%%%%%%%%%%%%%%%
%%%%%%%%%%%%%%%%%%%%%%%%%%%%%%%%%%%%%%%%%
There is now further compelling evidence that this unexpected pion mass dependence is a feature of QCD and not an artifact from the lattice.
At this time, we have no good idea of why this linear pion mass dependence arises.
But if we take it seriously, then the pion-nucleon sigma term would be
\begin{equation}
\sigma_{\pi N} = m_l \langle N | \bar{q}_l q_l | N \rangle 
	= m_l \frac{\partial}{\partial m_l} m_N(m_l)
	= 67 \pm 5 \textrm{ MeV}\, ,
	\qquad \textit{ruler approximation}\, .
\end{equation}
Recall, the sigma term is important for understanding the potential cross section of Dark Matter with a nucleus~\cite{Camalich:2013zqa}.  If $\sigma_{\pi N}$ is determined from fits to the nucleon mass from lattice QCD, one must be much more careful to account for finite volume and other systematic corrections, as this matrix element is determined from a derivative.

%Another important point to make concerns the chiral expansion of the nucleon mass.
At NLO in $\chi$PT, the nucleon mass is given by (ignoring the deltas)
\begin{equation}
m_N = M_0 + \sigma_N m_\pi^2
	- \frac{3\pi g_A^2}{(4\pi f_\pi)^2} m_\pi^3
	+ \mathcal{O}(m_\pi^4)\, ,
\end{equation}
where $\sigma_N m_\pi^2$ is the leading contribution to $\sigma_{\pi N}$.
The leading non-analytic light quark mass dependence to the nucleon mass is a prediction (if you take $g_A$ and $f_\pi$ from either phenomenology or LQCD calculations).
It would have been great (comforting, reassuring) if we were able to observe this non-analytic light quark mass dependence directly from LQCD calculations of $m_N$.  However, if one fits the $\chi$PT expression to the LQCD results, with $g_A$ as a free parameter, the resulting value of $g_A$ is quite small and inconsistent with the physical value.
In hindsight, this is rather obvious: given the observed pion mass dependence of the nucleon mass, there is no support for a contribution which scales as $(-m_\pi^3)$.
A fit using a value of $g_A$ from either nature or lattice QCD must include the $\mathcal{O}(m_\pi^4)$ terms for stability if extended to the range $m_\pi \sim 300$~MeV.

There have been a few attempts to understand the chiral behavior of $m_N$, for a recent example, see \cite{Alvarez-Ruso:2013fza}.  One, which I did not have time to discuss was work by M.~Alberg and G.~Miller~\cite{Alberg:2012wr}; using measured form-factors of the nucleon, the pion mass dependence of the nucleon was estimated by working on the light-front. 
A nearly linear dependence on the pion mass was found in the range where LQCD results exist.
Other attempts to understand the light quark mass dependence were motivated by the large-$N_c$ expansion.
In a first study~\cite{Jenkins:2009wv}, a comparison of the LQCD results of \cite{WalkerLoud:2008bp} were compared with the predicted scaling in $1/N_c$ and $SU(3)$ breaking, finding the predicted behavior~\cite{Jenkins:1995td} was satisfied for a large range of light quark masses with fixed strange quark mass.
The combined large $N_c$ and $SU(3)$ expansions predict certain linear combinations of the octet and decuplet baryon masses scale with definite powers of $N_c^{-1}$ and $m_s-m_l$.
A follow up study~\cite{WalkerLoud:2011ab} focussed on three linear combinations and the Gell-Mann--Okubo (GMO) relation
\begin{equation}
\begin{array}{l|l}
\textrm{mass relation}& \textrm{ scaling } \\
\hline
M_1 = 25(2m_N + m_\Lambda +3 m_\Sigma +2 m_\Xi)
	-4( 4m_\Delta +3m_{\Sigma^*} + 2m_{\Xi^*} + m_\Omega)
& N_c \times (m_s -m_l)^0
\\
M_3 = 5(6m_N + m_\Lambda -3m_\Sigma -4m_\Xi) -2(2m_\Delta -m_{\Xi^*} - m_\Omega)
& N_c^0 \times (m_s - m_l)
\\
M_4 = m_N +m_\Lambda -3m_\Sigma +m_\Xi
& N_c^{-1} \times (m_s - m_l)
\\
\Delta_{GMO} = \frac{3}{4}m_\Lambda +\frac{1}{4}m_\Sigma -\frac{1}{2} m_N -\frac{1}{2}m_\Xi
& N_c^{-1} \times (m_s - m_l)^{3/2}
\end{array}
\end{equation}
The most interesting result was that an $SU(3)$ baryon $\chi$PT fit to $M_3$ and $M_4$ returned values of the axial couplings consistent with phenomenology and direct LQCD calculations~\cite{Lin:2007ap},
\begin{equation}
D = 0.75(5)\, ,\quad 
F = 0.47(3)\, ,\quad
C = -1.4(1)\, ,\quad
H=-2.1(2)\, ,
\end{equation}
indirect evidence of non-analytic light quark mass dependence in the baryon spectrum.
Further, it was demonstrated that only the NNLO $SU(3)$ $\chi$PT formula was consistent with the LQCD results of the GMO relation.
This gives further evidence of non-analytic light quark mass dependence, as the leading contribution to the GMO relation is from non-analytic chiral loops.
The work was not definitive as there are known systematics which were not addressed.  Notably, the LQCD calculations were mixed-action with domain-wall valence quarks on the rooted-staggered MILC sea quark configurations, but the relevant mixed-action $\chi$PT~\cite{ma:eft} was not used:
the results exist at only a single lattice spacing:
volume corrections which are known to be more important for mass splittings~\cite{Beane:2011pc} were not addressed.
Despite these and other systematics, I want to stress the importance of finding baryon spectrum results consistent with values of the axial couplings known from both phenomenology and LQCD.

There has been additional interesting work comparing $SU(3)$ baryon $\chi$PT with LQCD results of the octet and decuplet spectrum.
There is compelling evidence that $SU(3)$ baryon $\chi$PT is a non-converging expansion~\cite{WalkerLoud:2008bp,baryons:su3}.%
%FOOTNOTE
\footnote{Refs.~\cite{WalkerLoud:2008bp,WalkerLoud:2011ab} explored the combined large $N_c$-$SU(3)$ baryon expansion which has better convergence behavior.}
However, in Refs.~\cite{lutz:su3}, reported by M.~Lutz here at CD12, $SU(3)$ $\chi$PT was used to fit the results of several different and independent LQCD calculations of the spectrum.
The most striking result is a consistent fit among all the LQCD results was found.  Further, the results were fit to a subset of the results, and then used to reasonably correctly ``predict" the results of other calculations, in some cases with very different values of the light and strange quark mass.
It is difficult to understand how this non-convergent expansion is used to so successfully describe all these different LQCD results.  There is something interesting to be learned from this.
It is worth noting, these findings have been independently confirmed by other groups as well~\cite{baryons:su3:more}.%
%FOOTNOTE
\footnote{
Another interesting method I did not have time to discuss is the new approach by the QCDSF Collaboration which varies the light and strange quark masses with the constraint $m_u + m_d + m_s = \textrm{ fixed}$~\cite{Bietenholz:2011qq}.
}

The last work I would like to bring attention to is that of T.~DeGrand who has been single handedly computing the baryon spectrum at different values of $N_c=3,5,7$, so far with quenched LQCD.  Good agreement between expected scaling with $N_c$ is found in the spectrum, including the splitting between different spin multiplets~\cite{DeGrand:2012hd}.

The advantage of considering the combined large $N_c$ and chiral expansions is both one of utility as well as one of formality.  The utility is that the large $N_c$ expansion, while not quantitative on its own, teaches us to ask better questions of our hadronic observables.  An example I gave above is considering not just the baryon spectrum, but linear combinations which scale with different powers of $1/N_c$.  In this way, a marginally convergent expansion can be improved with added factors of $1/N_c$, leading hopefully to quantitative improvements in our understanding.
There has been much debate in the literature on the ``correctness" of explicitly including the delta (decuplet) degrees of freedom in the low-energy effective field theory, see eg. previous Chiral Dynamics Workshops.
If one also incorporates the large $N_c$ expansion into $\chi$PT, this ambiguity is removed; in the $\lim_{N_c \rightarrow \infty}$, the nucleon and delta become degenerate.
This allows for a rigorous field-theoretic inclusion of these important degrees of freed in the Lagrangian.
Moreover, these large $N_c$ arguments help us understand certain phenomena which are only explicitly apparent when one explicitly includes the delta such as the nearly negligible pion mass dependence of $g_A$~\cite{gA_largeN} or the very small violations of the baryon GMO relation~\cite{Jenkins:1995gc}.
It is worth extending our understanding of these combined expansions and performing detailed comparisons with LQCD results.

%
%%
%%%
%%%%
%%%%%
\section{Baryons \textit{and} Lattice QCD}

A topic I have been particularly interested in is understanding the neutron-proton mass splitting from first principles.
As in my talk, I do not have enough room/time to discuss this topic in the detail I would like.  So here, I present a brief summary.

The neutron-proton mass splitting is known very precisely~\cite{Mohr:2008fa}
\begin{equation}
m_n - m_p = 1.29333217(42) \textrm{ MeV}\, ,
\end{equation}
yet our understanding of the separate electromagnetic and strong isospin breaking contributions is less than satisfactory.
This mass splitting plays a very significant role for example in Big Bang Nucleosynthesis (BBN): the initial ratio of neutrons to protons in the primordial universe is $X_n/X_p = \exp\{ -(m_n - m_p)/T\}$; the neutron lifetime is very sensitive to the mass splitting, changing approximately 100\% for a 10\% change in $m_n - m_p$.  For these reasons, among others, it is desirable to understand this important quantity from the fundamental theory.

At LO in isospin breaking, $m_n - m_p$ can be cleanly split into two pieces
\begin{equation}
m_n - m_p = \delta M_{n-p}^\gamma + \delta M_{n-p}^{m_d - m_u}\, .
\end{equation}
However, the quark mass operators are needed to renormalize the electromagnetic self-energy~\cite{Collins:1978hi} and so these contributions mix at higher orders.
Lattice QCD is a perfect tool to compute $\delta M_{n-p}^{m_d - m_u}$, for which there are a few existing results~\cite{mn-mp:md-mu}.
Computing the electromagnetic contribution is much more challenging with LQCD due to the disparate length scales relevant for QCD and QED.  With C.~Carlson and G.~Miller, I revisited the Cottingham~\cite{Cottingham:1963zz} formulation of determining the electromagnetic self-energy using dispersion relations to relate the measure Compton cross sections to the self-energy~\cite{WalkerLoud:2012bg}.
This work uncovered a technical oversight in work of Gasser and Leutwyler~\cite{Gasser:1974wd} related to a subtracted dispersion integral, which unfortunately invalidates their result.
After properly accounting for the subtraction function, the electromagnetic self-energy contribution to $m_n-m_p$ can be broken into four pieces
\begin{equation}\label{eq:dMg_terms}
 \d M_{p-n}^\gamma = \d M^{el} + \d M^{inel} + \d M^{sub} + \d \tilde{M}^{ct}\, .
\end{equation}
One can show the residual counterterm contribution (after renormalization) is numerically second order in isospin breaking and can be neglected~\cite{Collins:1978hi}, $|\d \tilde{M}^{ct}| \lesssim 0.02$~MeV~\cite{WalkerLoud:2012bg}.
The elastic ($\d M^{el}$) and inelastic ($\d M^{inel}$) contributions can be precisely computed, but the unknown subtraction function presents challenges.
The $\lim_{Q^2\rightarrow \infty}$ behavior of the subtraction function can be shown to scale as $1/Q^2$~\cite{Collins:1978hi} while the low-energy limit can be constrained with effective field theory~\cite{Pachucki:1996zza}.
Introducing a form factor to connect the low and high $Q^2$ regions, the unknown subtraction function can be modeled as
\begin{equation}\label{eq:sub_el_inel}
\d M^{sub}_{inel} \simeq \frac{-3 \beta_M^{p-n}}{8\pi} \int_0^{\L^2} {\hskip-0.6em} dQ^2 
	Q^2 \left( \frac{m_0^2}{m_0^2 + Q^2} \right)^2 {\hskip-0.4em}\, ,
\end{equation}
where $m_0$ can be taken as a typical dipole mass and $\beta_M^{p-n}$ is the isovector magnetic polarizability.
Unfortunately, this polarizability is presently unknown at the 100\% level~\cite{Griesshammer:2012we}, and so at best we can estimate this contribution;
$\d M^{sub}_{inel} = 0.47 \pm 0.47 \textrm{ MeV}$.
Putting all the contributions together,
\begin{equation}
	\delta M_{n-p}^\gamma = -1.30 \pm 0.03 \pm 0.47 \textrm{ MeV}\, ,
\end{equation}
where the first uncertainty is from the elastic and inelastic contributions and the second, from the unknown subtraction function.
This value is in good agreement with expectations from combining LQCD~\cite{mn-mp:md-mu} and experiment~\cite{Mohr:2008fa}, which would lead one to predict $\delta M_{n-p}^\gamma = -1.42(15)$~MeV.
LQCD calculations of the isovector magnetic polarizability will improve this prediction, and allow for a more meaningful comparison with other lattice results and more importantly, with nature.

%
%%
%%%
%%%%
%%%%%
\section{Conclusions}
This is a particularly exciting time.  Lattice QCD calculations are now being performed with light quark masses at or near their physical values, opening a new opportunity to constrain $\chi$PT and determine  the LECs appearing in the chiral expansion.
In the last five years, we have witnessed the utility of LQCD in constraining High Energy Physics phenomenology to improve our understanding of the Standard Model and aid in the search for new physics.
There is now a working group tasked with computing lattice averages of basic quantities of interest for light quark meson physics~\cite{Colangelo:2010et}.
In the next five years, we will see LQCD results transform our understanding of low energy Chiral Dynamics by solidifying the connection between nuclear physics and QCD.
Hopefully, along the way, we will uncover and solve new puzzles, beginning with an understanding of the nucleon.

\end{document}